\documentclass[a4paper]{spie}  
\usepackage[]{graphicx}

\usepackage{epsfig}
\usepackage{amssymb}
\usepackage{amsmath}
\usepackage[english,francais]{babel}
\usepackage[T1]{fontenc}

\title{Apodized Lyot Coronagraph for VLT-SPHERE\,: Laboratory tests and performances of a first prototype in the visible} 

\author{Géraldine Guerri\supit{a}, Sylvie Robbe-Dubois\supit{a}, Jean-Baptiste Daban\supit{a}, Lyu Abe\supit{a}, Richard Douet\supit{a},
Philippe Bendjoya\supit{a}, Farrokh Vakili\supit{a}, Marcel Carbillet\supit{a}, Jean-Luc Beuzit\supit{b}, Pascal Puget\supit{b}, Kjetil Dohlen\supit{c} and David Mouillet\supit{b} \skiplinehalf
\supit{a}Laboratoire A.H. Fizeau, UMR-CNRS 6525, Universit\'{e} de Nice Sophia-Antipolis, Observatoire de la C\^{o}te d'Azur, Parc Valrose, 06108 Nice Cedex 2, France\\
\supit{b}Laboratoire d'Astrophysique Observatoire de Grenoble (LAOG), UMR-CNRS 5571, Universit\'{e} J. Fourier, 414, rue de la piscine, 38041 Grenoble cedex 9, France\\
\supit{c}Laboratoire d'Astrophysique de Marseille (LAM), Observatoire Astronomique de Marseille Provence, 2, place Le Verrier, 13248 Marseille cedex 4, France}

\authorinfo{Further author information: Send correspondence to G. Guerri, guerri@unice.fr, Telephone:  +33 (0)4 92 07 65 73 }

\begin{document} 
  \maketitle 

\begin{abstract}
We present some of the High Dynamic Range Imaging activities developed around the coronagraphic test-bench of the Laboratoire A. H. Fizeau (Nice).
They concern research and development of an Apodized Lyot Coronagraph (ALC) for the VLT-SPHERE instrument and experimental results from our testbed working in the visible domain.

We determined by numerical simulations the specifications of the apodizing filter 
and searched the best technological process to manufacture it. 
We present the results of the experimental tests on the first apodizer prototype in the visible and the resulting ALC nulling performances.

The tests concern particularly the apodizer characterization (average transmission radial profile, global reflectivity and transmittivity in the visible), ALC nulling performances compared with expectations, sensitivity of the ALC  performances to misalignments of its components. 


\end{abstract}



\section{INTRODUCTION}
\label{sec:intro}  
Since 1995 and the discovery of the first extrasolar planet by M. Mayor and D. Queloz (1995~\cite{Mayor95}), direct detection
and spectral characterization of an exoplanet has become one of the most exciting and challenging astromical areas.
In this context, European Southern Observatory (ESO) supported two phase A studies for a "planet finder" instrument for its second-generation instruments on the Very Large Telescope (VLT).
After the review of these two studies, an unique instrument called SPHERE (Spectro-Polarimetric High-contrast Exoplanet REsearch) 
is now considered for a first light in 2010.

SPHERE (Beuzit et al. 2006~\cite{Beuzit06}) is a second generation instrument for the VLT designed and built by a consortium of French, German, Italian, Swiss and Dutch 
institutes in collaboration with ESO. The project is currently in its phase B.
The main goal of SPHERE is the direct detection of faint objects very close to a bright star, especially giant extrasolar planets. Other science studies concern brown dwarfs, circumstellar disks and related phenomena such as mass loss mechanisms, stellar winds or planetary nebulae.
The design of the SPHERE instrument is divided into four subsystems\,: the Common Path Optics and three science channels, a differential imaging camera (IRDIS), an Integral Field Spectrograph (IFS), and a visible imaging polarimeter (ZIMPOL).

In the consortium, we are in charge of the study and the development of an Apodized Lyot Coronagraph (ALC). In the instrument design, 
the ALC will be a part of the Common Path Optics subsystem of SPHERE.

The purpose of this paper is twofold. 
After having briefly recalled the principle of the Apodized Lyot Coronagraph in section 2, we describe in section 3 the experimental setup of the high dynamics range imaging coronagraphic testbed and the detailed characteristics of the three main components of the ALC tested prototype (apodizer, coronagraphic mask and Lyot stop). 
Then, we present the laboratory tests of the ALC prototype in the visible. They concern measurements of the apodizer transmission profile (section 4) and of the ALC coronagraphic performances (section 5) and the study of the ALC sensitivity to lateral and longitudinal misalignment of its components (section 6).   

Although SPHERE will operate in the near-IR domain, these visible measurements allowed to validate the principle and the manufacturing technique of the apodizer.
It gave preliminary estimations of the coronagraph performance. Furthermore, a visible optical bench is easier to set-up than a near IR cold bench.

\section{PRINCIPLE OF THE APODIZED LYOT CORONAGRAPH} 
\label{sec:ALC_Principle}

\subsection{Preliminary}
Firstly, Bernard Lyot introduced in 1930 the principle of the Classical Lyot Coronagraph (Lyot, 1930~\cite{Lyot30})\,: an occulting disk, the Lyot coronagraphic mask, is placed in the telescope focal plane so as to block the central part of the Airy pattern of the star.
In the relayed pupil plane, a diaphragm called Lyot stop is placed in order to remove the light rejected by the coronagraphic mask out of the geometrical image of the pupil. The smaller the size of this Lyot stop is, the better the light attenuation is, but also the lower the throughput is.
Unfortunately, the contrast achieved by the Classical Lyot Coronagraph is not sufficient enough to image an exoplanet.
%

\subsection{Principle}
\begin{figure}[!h]
  \centering
 \includegraphics[clip=true,width=0.6\columnwidth]{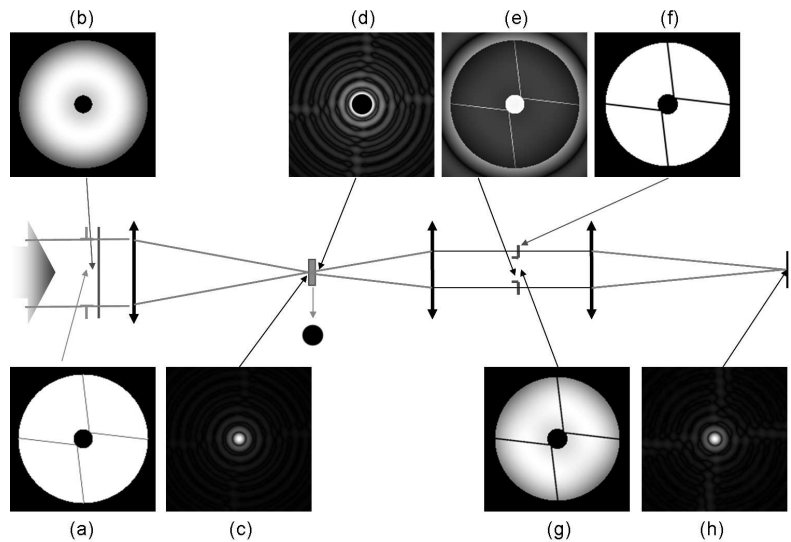}
  \caption{Principle of SPHERE Apodized Lyot Coronagraph\,: (a) Entrance pupil, (b) Apodizer, (c) Point spread function (PSF) at the focus of the telescope, (d) PSF when the Lyot occulting coronagraphic mask is settled, (e) Pupil image before the Lyot stop introduction, (f) Lyot stop, 
(g) Pupil image with the Lyot stop, (h) Final coronagraphic PSF}
  \label{fig:Princ_ALC}
\end{figure}

In this context, Aime, Soummer et al. (2002 \cite{Aime02} , 2003 \cite{Soummer03}) proposed a dramatic improvement with the principle of the Prolate Apodized Lyot Coronagraph.
In Fig.~\ref{fig:Princ_ALC}, we present the principle of the Apodized Lyot Coronagraph (ALC) 
adapted to our study, ie the ALC designed for SPHERE.

In this figure, we can see that, because the VLT entrance pupil is not a simple circular aperture and has a central obscuration and 4 spiders, the apodizer transmission function has a bagel shape (Soummer 2005\cite{Soummer05} , Carbillet et al. in prep.\cite{Carbillet08}).
Furthermore, the best coronagraphic performance is obtained when the Lyot stop has the same shape as the entrance pupil one.

%

\subsection{Overview of apodizer manufacturing techniques}
Apodizing masks are commonly used in optics for 20 years for instance for laser beam shaping.
However, the shape of the transmission profiles and the tolerance bounds are totally different and less constraining than those needed to perform astronomical high dynamic range imaging.
That's why, one of the ALC critical realization point is the manufacturing of the apodizer. Several worldwide institutes are currently testing various apodization techniques\,: we can cite for instance thin layer deposition, ion implantation, HEBS\texttrademark 
glass (Soummer et al. 2006\cite{Soummer06}) or use of a Mach-Zehnder interferometer (Carlotti et al. 2007\cite{Carlotti07}).
The main challenge is to obtain an apodizer that meets as much as possible the strong specifications and who introduces
the lowest wavefront errors.



\section{EXPERIMENTAL SETUP OF THE CORONAGRAPHIC BENCH} 
\label{sec:Exp_Setup}

\subsection{The coronagraphic bench} 
\label{ssec:Corono_bench}
\begin{figure} [!h]
   \begin{center}   
  \includegraphics[height=5cm,width=0.96\linewidth]{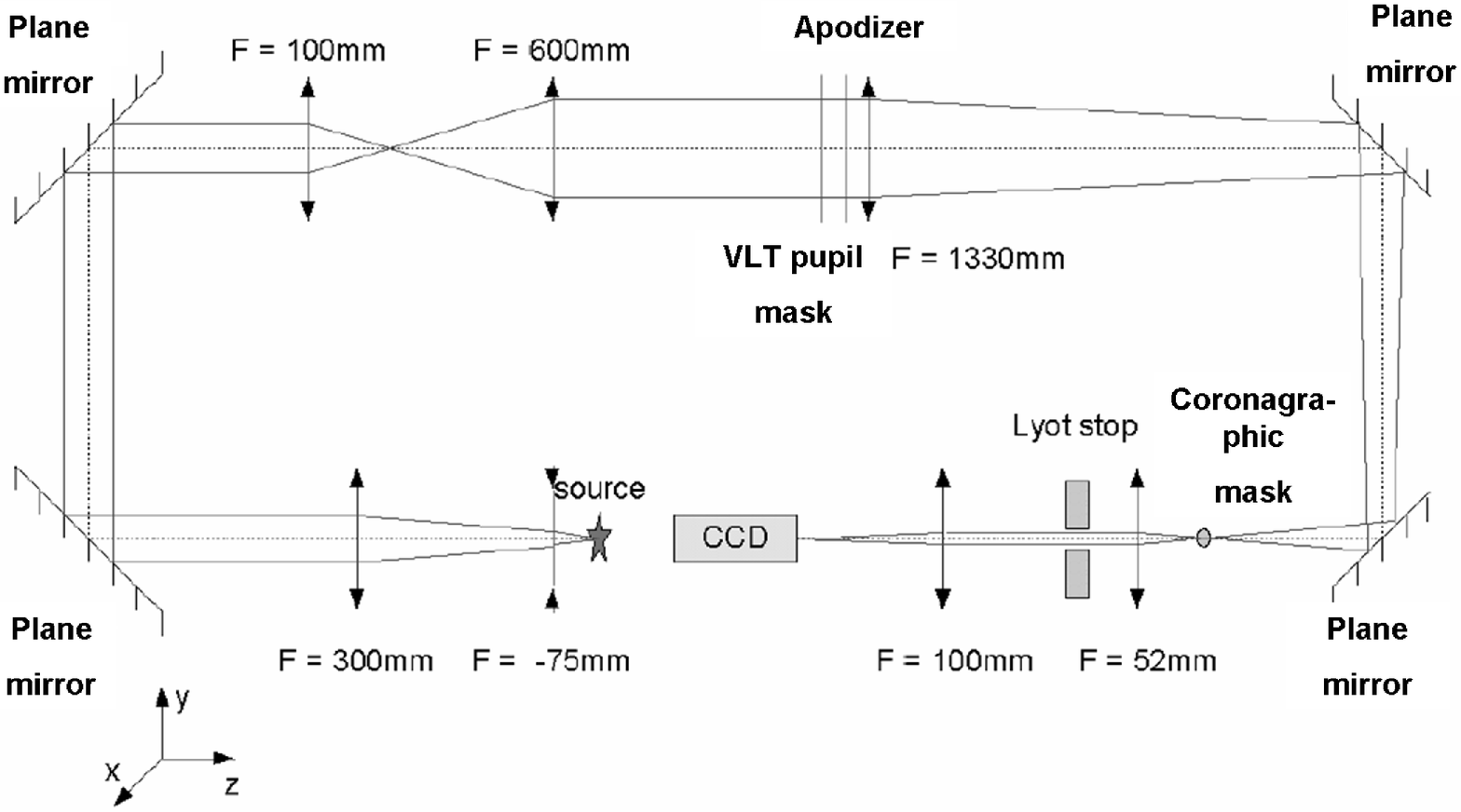}  
  \caption{Optical setup of Fizeau laboratory high dynamics range imaging bench.}
  \label{fig:Setup_Bench}
	\end{center}
   \end{figure}

Figure~\ref{fig:Setup_Bench} shows the visible optical setup of the high dynamics range imaging bench developed by the Fizeau laboratory.
The characteristics of each optical element are given in table~\ref{tab:Descriptif_Composants_Banc_ITHD}.

\noindent Several light sources are available\,:
\begin{itemize}
\item a HeNe-laser Melles Griot 25 LHR 691-230\,: $\lambda=632.8~nm$, output power 2.5mW for alignment purpose. 
\item a fibered laser diode Melles Griot 57PNL062/P4\,: $\lambda=635~nm~\pm 15$, output power 7mW, polarization-maintaining singlemode fiber, for monochromatic performance (less coherent than the laser). 
\item a white light source with an output power of 5mW for chromatic measurements.
\end{itemize}
\begin{table} [!h]
\begin{small}
\begin{center}
\begin{tabular}{|c|c|}
\hline
Component                           					         					& Characteristics \\
\hline\hline
 Fibered laser diode                                        				& $\lambda$ = 635 nm, P = 7 mW\\
 Diverging lens  $L_{1}$                           				&    $f_{1}$ = -75 mm\\
 Converging lens  $L_{2}$                        				&    $f_{2}$ = +300 mm\\
 Plane mirrors $M_{1}$ and $M_{2}$                	& optical quality $\lambda$/5~at~633nm \\
 Converging lens $L_{3}$                       				& $f_{3}$ = +100 mm \\
 Converging lens $L_{4}$                       				& $f_{4}$ = +600 mm \\
 Pupil mask                                       					& see Fig~\ref{fig:Mask_Pup_VLT} and Sect.~\ref{sssec:Mask_Pup}\\
 Apodizer        										    			& see Fig~\ref{fig:Spec_Trans_Apo_3Lsd} and Sect.~\ref{sssec:Apo}\\
 Converging lens $L_{5}$ (achromat)    				& $f_{5}$ = +1330 mm \\
 Miroirs plans $M_{3}$ and $M_{4}$               				& optical quality $\lambda$/5~at~633nm\\
 Lyot coronagraphic mask              				& $\Phi$= $90~\mu$m see Sect.~\ref{sssec:Coronomask}\\
 Converging relay lens L$_{6}$ (achromat)             &$f_{6}$ = +52 mm\\
 Lyot stop                								    						& see Sect.~\ref{sssec:LS}  \\
 Imaging converging lens L$_{7}$ (achromat)    &$f_{7}$ = +100mm \\
 Science camera  					            					& CCD Adimec\\
\hline
\end{tabular}
\end{center}
\end{small}
\caption{Inventory of the Fizeau laboratory high dynamics range imaging bench components.}
\label{tab:Descriptif_Composants_Banc_ITHD}
\end{table}
%

The lenses $L_{1}$, $L_{2}$, $L_{3}$ et $L_{4}$ form two successive afocal systems that collimate the beam in order to obtain a parallel beam
behind the lens $L_{4}$. The diameter of the beam is here $28~mm$, equal to the apodizer one. 
This resulting beam was thereby magnified by a factor of 24 compared to the initial beam so as to keep the central part of the gaussian wavefront 
delivered by the fibered source and therefore to obtain an uniform intensity distribution in the pupil plane.
Nevertheless the size of the useful part remains $28~mm$.   
Then comes the so called coronagraphic chain whose components will be described below. 
It should be noted that the four plane mirrors have the unique fonction to fold up the beam in the available location.       
The detector is a CCD array Adimec-1000M.
The main characteristics of this detector are\,: array of 1004x1004 pixels (7.4 $\mu$m each), 10 bits dynamics, 16 e- read-out noise.

%
%
\subsection{Characteristics of the ALC components} 
\label{ssec: }

\subsubsection{The pupil mask}
\label{sssec:Mask_Pup} 
\begin{figure} [!h]
   \begin{center}   
\begin{tabular}{cc}
 \includegraphics[width=0.3\linewidth]{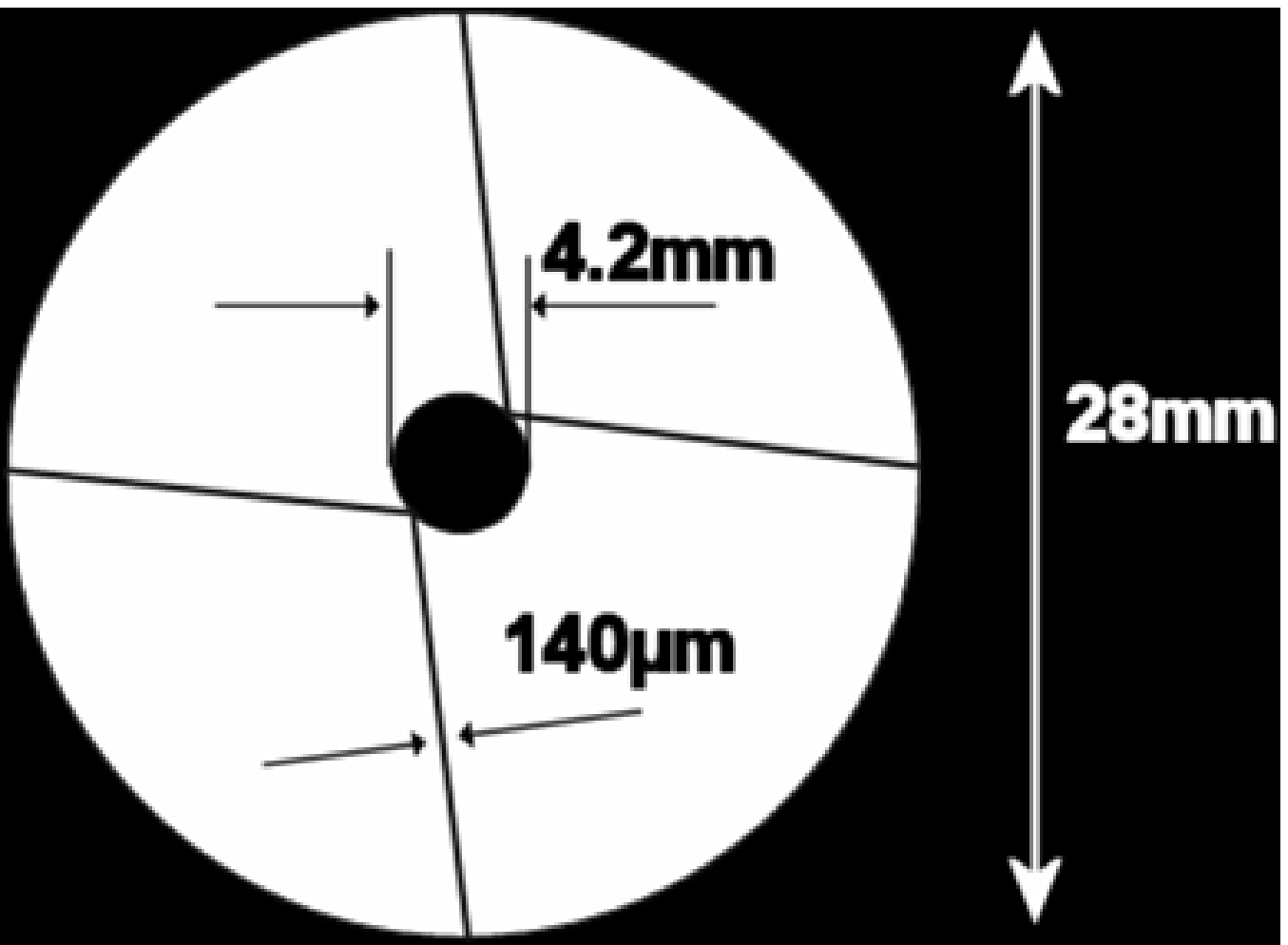}  &
  \includegraphics[width=0.3\linewidth]{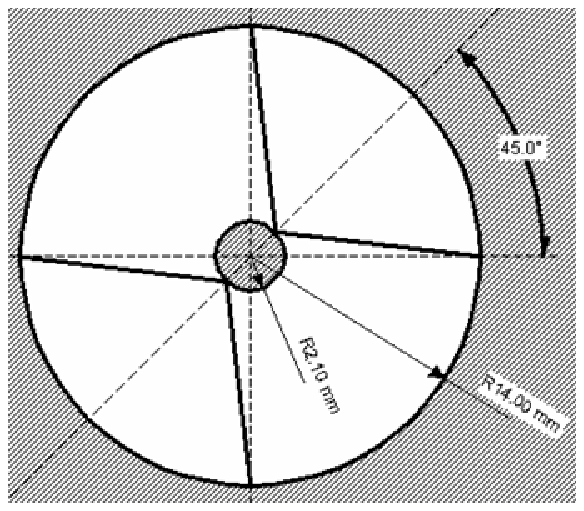}  \\
\end{tabular}

  \caption{Shape and dimensions of the pupil mask reproducing the VLT entrance pupil.}
  \label{fig:Mask_Pup_VLT}
	\end{center}
   \end{figure}

The pupil mask reproduces the VLT entrance pupil as shown on figure~\ref{fig:Mask_Pup_VLT}. It was manufactured by the company Micromodule (Brest, France) by a Chromium deposition on a BK7 substrate.
The mask is 100\% transmissive on a  $28~mm$-diameter disk called outer diameter. The mask blocks the light on a concentric central disk that simulates the telescope central obscuration (its $4.2~mm$ diameter corresponds to 15\% of the outer diameter).
Four rectilinear arms simulate the telescope spiders, $140~\mu$m-width each. They join the four cardinal points of the outer diameter to the central obsruction in 2 fastening points. 
Table~\ref{tab:Spec_MaskPup} gives the technical specifications of the pupil mask.
\begin{table} [!h]
\begin{small}
\begin{center}
\begin{tabular}{|c|c|}
\hline
Parameter                          			     & Specification  \\
\hline\hline
Outer diameter                             & $28.00~\pm 0.05~mm$  \\
Central obscuration diameter         & $4.20~\pm 0.05~mm$ \\
Spiders width                         & $0.140~\pm 0.01~mm$  \\
Mask optical density                            & $10^{6}~\pm 10^{0.5}$  \\
Substrate                                 &BK7  \\
Substrate thickness                         &$1.5~mm$ \\
Substrate optical quality          & $\lambda/4$ PTV at $\lambda=633~nm$ on the front face  \\
Substrate parallelism                       & $\leq 1'$  \\
Antireflection coating                             & on the back face (<0.5\%)  \\
\hline
\end{tabular}
\end{center}
\end{small}
\caption{Technical specifications of the pupil mask.}
\label{tab:Spec_MaskPup}
\end{table}
%
%
\subsubsection{The apodizer}
\label{sssec:Apo} 
The apodizer was manufactured by the company Reynard Corporation according to the metallic thin layer evaporation technique
of Inconel 600\texttrademark ~on a BK7 substrate.
Its transmission profile was optimized by numerical simulations to fit with a Lyot coronagraphic mask with an angular diameter of $3\lambda/D$. 
Table~\ref{tab:Spec_Apo} gives the technical specifications for the apodizer that were required to the manufacturer.
\begin{table} [!h]
\begin{small}
\begin{center}
\begin{tabular}{|c|c|}
\hline
Characteristics                           			         & Specification  \\
\hline\hline
Outer diameter of the apodized area            &  $28.00~mm$ \\
Outer diameter of the substrate                         &  $40.00~mm$ \\
Substrate thickness                                        &  $2~mm$ \\
Surface quality of the substrate (scratch - dig)   &  $5$ - $10$     \\
Optical surface quality                                & $\lambda /10$ PTV at 632.8 nm    \\
Parallelism                                                    & $\leq~1'$       \\  
Substrate material                                     & BK7  \\
Coated metal                                                & Inconel  \\
Tolerance on the profile                                & $\pm 5 \%$  \\
Coating                                                       & Anti-reflection coating on the 2 faces \\
\hline
\end{tabular}
\end{center}
\end{small}
\caption{Manufacturing specifications for the apodizer 3$\lambda$/D in the visible.}
\label{tab:Spec_Apo}
\end{table}

\par
\begin{figure} [!h]
   \begin{center}   
   \includegraphics[width=.7\linewidth]{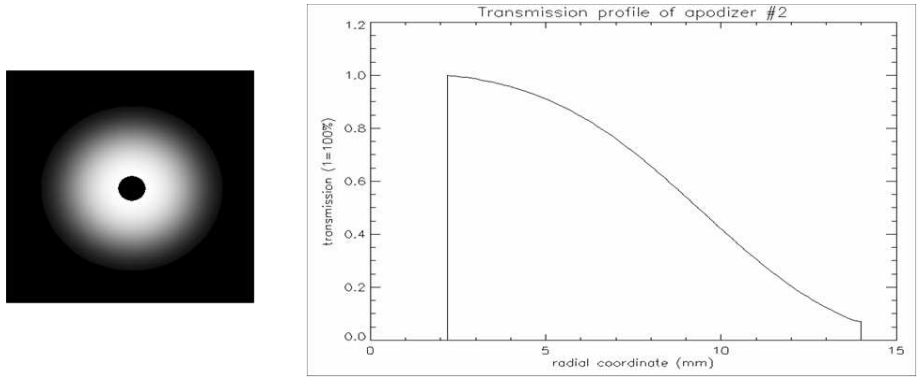}  
      \caption{Theoretical transmission of the apodizer in the visible send to the manufacturer\,: (left) 2D view, (right) optimal average radiale profile in transmission.}
   \label{fig:Spec_Trans_Apo_3Lsd}
   \end{center}
\end{figure}

\noindent Figure~\ref{fig:Spec_Trans_Apo_3Lsd} shows the theoretical optimal average radial profile in transmission of the apodizer.
%
This profile, noted $T(r,\theta)$, can be be fitted from the numerical simulations. It is approximated by the 7\textsuperscript{th} order polynomial defined by\,:
\begin{displaymath}
\forall \theta ~~ T(r,\theta)~=~\sum_{k = 0}^{7}  C_{k}*(\dfrac{r}{14})^{k}~with~:
\end{displaymath}
%
%
\begin{itemize}
  \item 	r given in mm and $2.1\leqslant r \leqslant 14$.
  \item 	$C_{0}	 = 0.95610110$,
$C_{1}	 = 0.99129289$,
$C_{2}	 = - 7.6010470$,			
$C_{3}	 = 26.988797$,	
$C_{4}	 = - 61.164585$,	
$C_{5}	 = 70.131699$,	
$C_{6}	 = - 38.251457$ and	
$C_{7}	 = 8.0134063$.	
 \item The transmission profile in the area of radius smaller than $2.1~mm$ is of a lesser importance because this area will be occulted by the central obscuration of the pupil mask.
\end{itemize}	
\vspace{0.25cm}
The complete experimental characterization of the apodizer is presented in the next section. 
%
%
\subsubsection{The Lyot coronagraphic mask}
\label{sssec:Coronomask} 

The Lyot coronagraphic mask blocks the light on a central disk of $90~\mu m$ diameter (which corresponds to an angular diameter of 3$\lambda$/D) 
and is transparent outside.
This element was also manufactured by the company Micromodule by a Chromium deposition on a BK7-substrate. Simulations allowed to determine that a value of 6 for the optical density of the opaque zone is necessary to obtain the desired coronagraphic effect.
Table~\ref{tab:Spec_Coronomask} gives the manufacturing specifications of the coronagraphic mask.
\begin{table} [!h]
\begin{small}
\begin{center}
\begin{tabular}{|c|c|}
\hline
Characteristics                           			                     & Specification  \\
\hline\hline
Tolerance on the component width (square)                  & $15.0~\pm 1.0~mm$  \\
Tolerance on the coronagraphic mask diameter       & $90.0~\pm 1.0~mm$  \\
Mask optical density                                           & $6.0~\pm 0.5$  \\
Substrate type                                                          & BK7 or N-BK7  \\
Substrate thickness                                                  &  $1.5~mm$ \\
Substrate optical quality                                           & $\lambda/4$ PTV at $\lambda=633~nm$ on the front face   \\
Substrate parallelism                                               & $\leq 1'$  \\
\hline
\end{tabular}
\end{center}
\end{small}
\caption{Specifications of the 3$\lambda$/D Lyot coronagraphic mask.}
\label{tab:Spec_Coronomask}
\end{table}
%
%
\subsubsection{The Lyot stop}
\label{sssec:LS} 

The Lyot stop has the same shape as the one of the VLT pupil mask. It was also manufactured by the Micromodule company by a Chromium deposition on a BK7 substrate. 
This diaphragm is geometrically homothetical to the pupil mask\,: the central obscuration diameter is slightly enlarged, the diameter of the external disk is slightly reduced, the spider arms are magnified but their orientation is identical to the pupil mask ones.

Some simulations showed that the best Lyot diaphragm (according to a criterion obtaining the optimal attenuation without too reducing the transmission) has the following properties : an outer diameter corresponding to 97\% of the size of the pupil image, a central obscuration that is 1.05 times larger compared to the obstruction size in the coronagraphic pupil plane, and spiders arms magnified by a factor 2. 
Furthermore, experience inherited from previous laboratory tests showed that it is preferable to slightly undersize the Lyot stop size to 
get rid of some instrumental biases such as, for instance, egde diffraction effects or tolerances on the lenses focal lengths. 

Table~\ref{tab:Spec_LS1} gives the manufacturing specifications of the Lyot stop.

\begin{table} [!h]
\begin{small}
\begin{center}
\begin{tabular}{|c|c|}
\hline
Parameter                          			     & Specification  \\
\hline\hline
Outer diameter                               & $1.060~\pm 0.01~mm$  \\
Central obscuration diameter         & $0.170~\pm 0.01~mm$ \\
Spiders width                                  & $0.010~\pm 0.002~mm$  \\
Mask optical density                       & $6~\pm 0.5$  \\
Type de substrat                            &BK7  \\
Substrate thickness                        &  $2~mm$ \\
Substrate optical quality                 & $\lambda/4$ PTV at $\lambda=633~nm$ on the front face \\
Substrate parallelism                      & $\leq 1'$  \\
\hline
\end{tabular}
\end{center}
\end{small}
\caption{Lyot stop manufacturing specifications.}
\label{tab:Spec_LS1}
\end{table}
%


\section{APODIZER EXPERIMENTAL CHARACTERIZATION} 
\label{sec: }

\subsection{Apodizer average transmission profile}
\label{ssec:Trans_Apo_HeNe}

\par\noindent\textbf{Principle.}
This measurement is done with the white light source coupled with a narrowband interferential filter ($\lambda_{c}=603~nm$, $\delta \lambda=25~nm$) placed in front of the source.
The principle of the measurement is to acquire pupil images with and without apodizer, called respectively apodized and reference images. 
After a dark substraction, a long exposure is generated for both pupil image types. 
Then, the apodized long exposure is divided by the reference one in order to avoid the inhomogenous illumination of the reference pupil. 
The radial transmission values of the apodizer are finally computed from this divided long exposure image. 

\vspace{0.25cm}
\par
\begin{figure} [!h]
   \begin{center}   
\parbox{3cm}{ (a) \includegraphics[height=2.5cm]{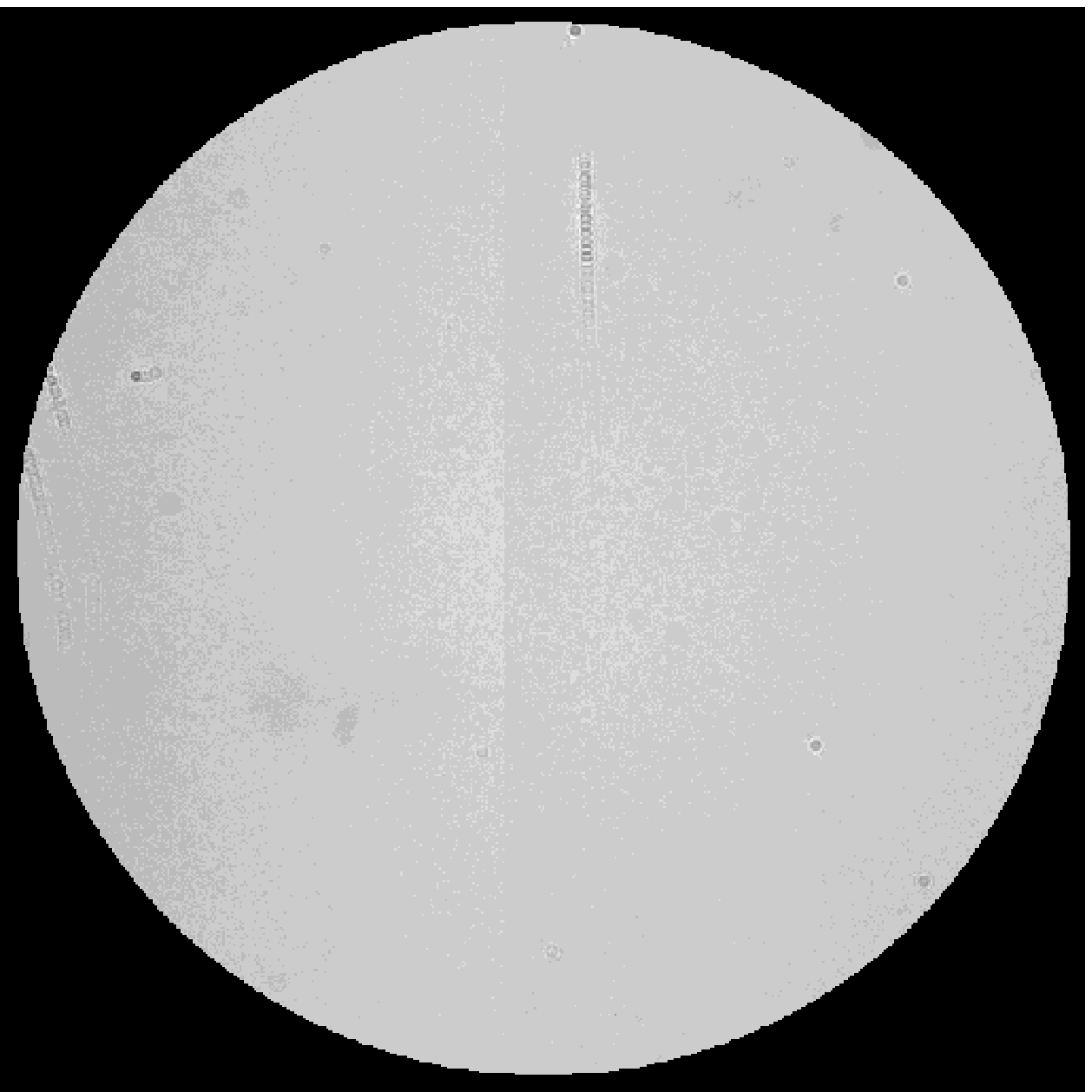}  }
\parbox {3cm}{ (b) \includegraphics[height=2.5cm]{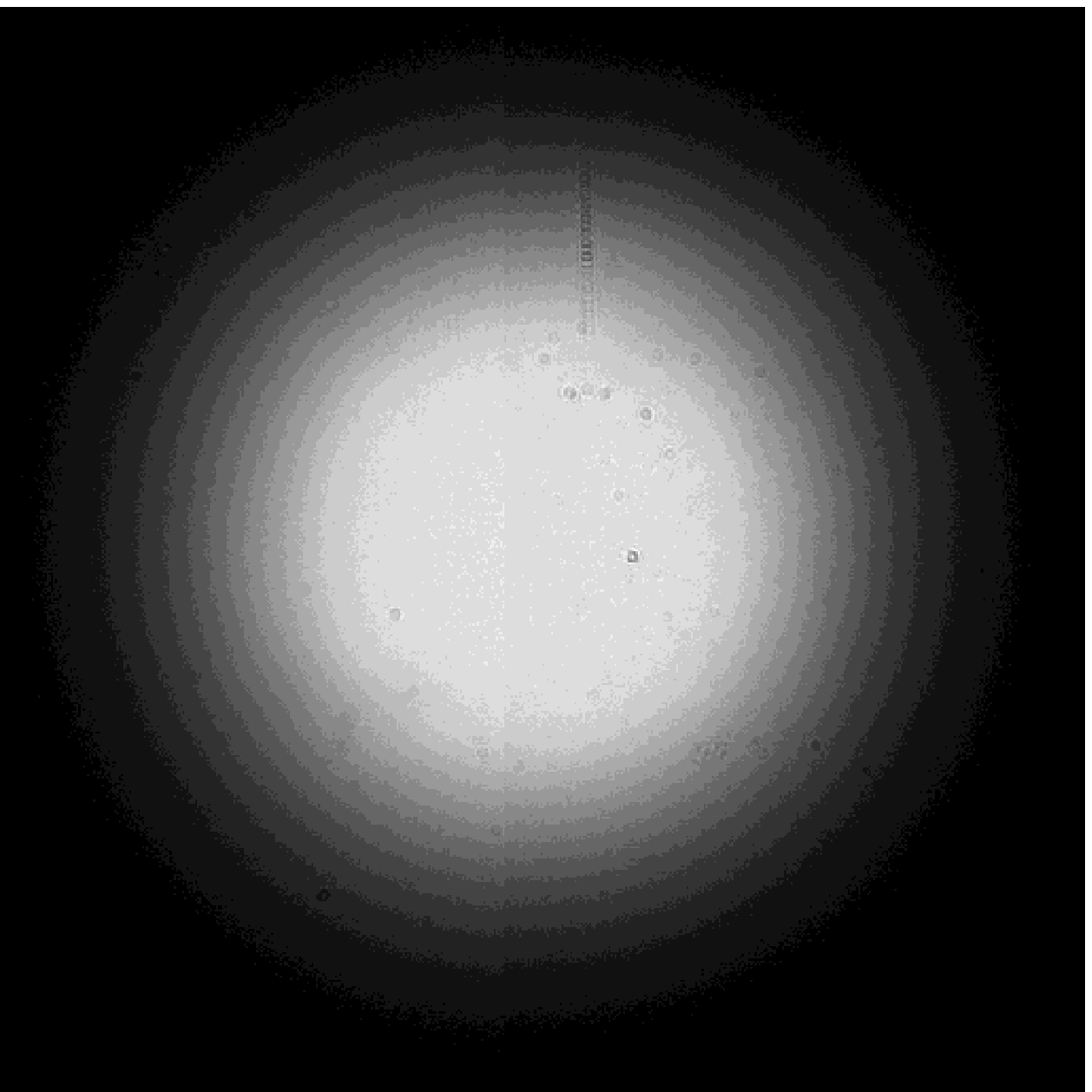} }
\parbox {6.5cm}{ (c) \includegraphics[height=6cm]{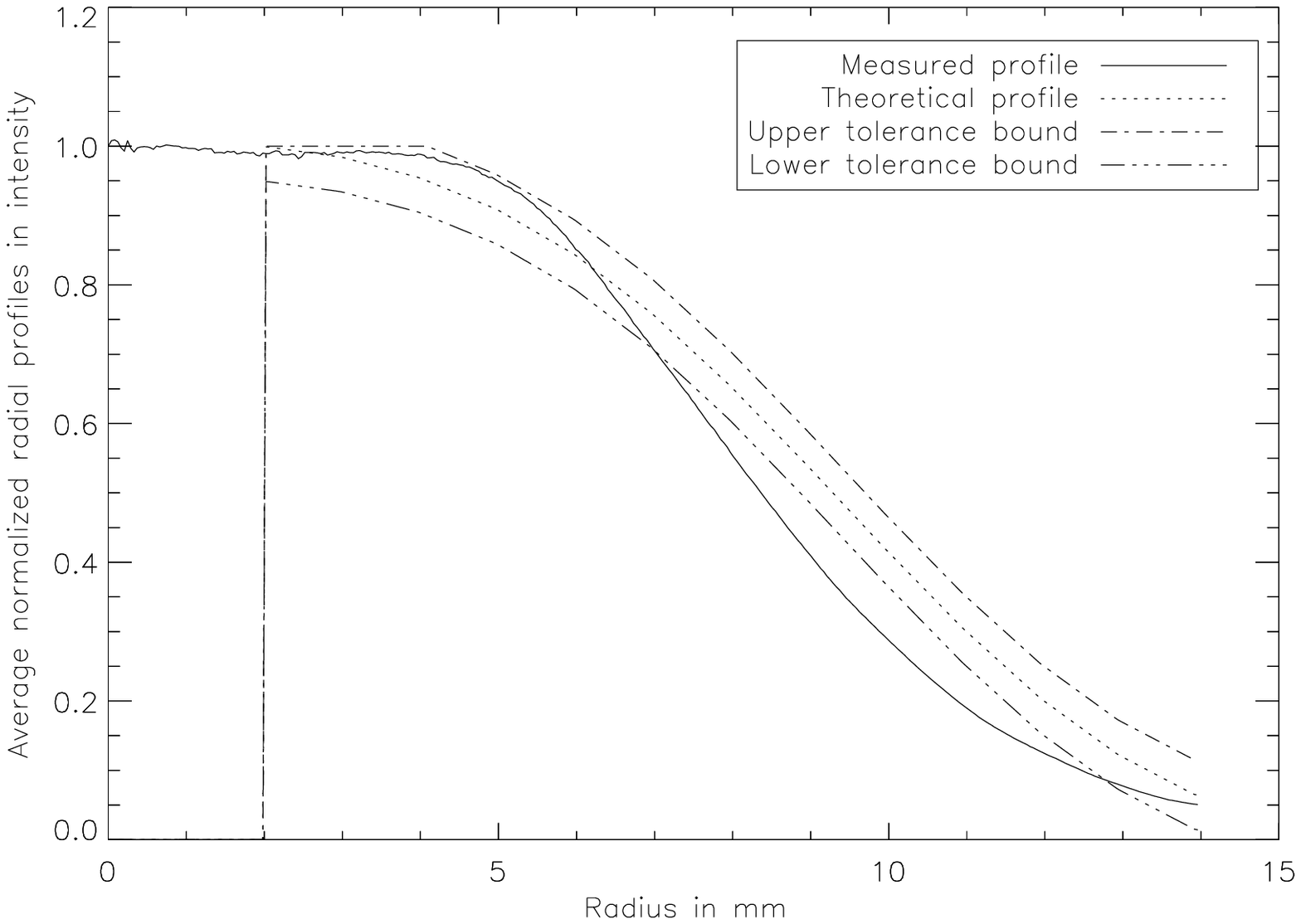}  }
 \caption{2D-views of (a) the reference and of (b) the apodised pupils, (c) average radial transmission profile of the apodizer\,: measured profile, theoretical profile, tolerance bounds.}
  \label{fig:Mes_Trans_Apo_HeNe2}
	\end{center}
   \end{figure}

\noindent\textbf{Results.}
Fig.~\ref{fig:Mes_Trans_Apo_HeNe2} shows an example of a reference and an apodized long exposure pupil images. Both long exposures are obtained by co-adding 5 snapshots of a 2 ms exposition time. 
It is then necessary to compare the measurement to the theoretical profile and to the tolerance limits\,: figure~\ref{fig:Mes_Trans_Apo_HeNe2} shows the different average radial profiles.

\noindent The measurement doesn't meet the specifications in the area where r is included between $7$ mm and $12.8$ mm\,: the lower tolerance limit is overrun.
Besides, the apodizer global transmission coefficient worths 39.9\%, as expected in the specifications.
%
%
\subsection{Apodizer global transmittivity in the visible}
\label{ssec:}

To realize this measurement, we use the same experimental protocol as in the
last section, using 15 spectral filters.
Table~\ref{tab:Descriptif_Filtres_Spectraux} lists the properties of the 15 available spectral filters.
\begin{table} [!h]
\begin{scriptsize}
\begin{center}
\begin{tabular}{|c|c|c|c|c|c|c|c|c|c|c|c|c|c|c|c|}

\hline
Number&1&2&3&4&5&6&7&8&9&10&11&12&13&14&15\\
\hline\hline
$\lambda_{c} $ (nm) &421.5 &432.5 &445 &485 &513.5 &535.5 &550 &545 &558 &563.5 &574 &584.5 &603.5 &661 &700.5\\
$\lambda_{min}$ (nm) &412  & 425& 433 & 475 & 502 & 525 & 535 & 535 & 547 & 552 & 565 & 569 & 587 & 652 & 689 \\
$\lambda_{max}$ (nm) & 431 & 440 & 457 & 495 & 525 & 546 & 565 & 555 & 569 & 575 & 583 & 600 & 620 & 670 & 712\\
$\Delta \lambda$ (nm) & 19 & 15 & 24 & 20 & 23 & 21 & 30 & 20 & 22  & 23 & 18 & 31 & 33 & 28 & 23\\
\hline

\end{tabular}
\end{center}
\end{scriptsize}
    \caption{Inventory of the spectral filters.}
    \label{tab:Descriptif_Filtres_Spectraux}
\end{table}

\begin{figure} [!h]
   \begin{center}   
  \includegraphics[height=5cm]{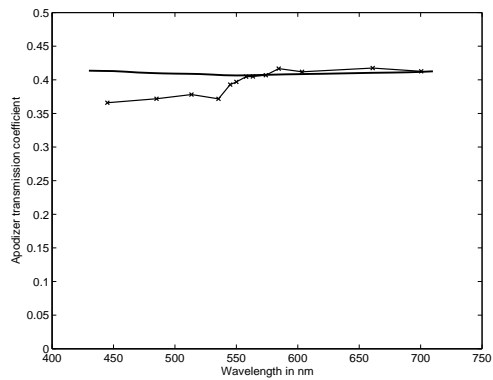} 
  \caption{Measurement of the transmission coefficient of the apodizer (crosses) compared to the numerical simulation (solid line).}
  \label{fig:Mes_Trans_Apo_3Lsd}
	\end{center}
	\vspace*{-0.25cm}
   \end{figure}

Figure~\ref{fig:Mes_Trans_Apo_3Lsd} shows the spectral evolution of the global transmission of the apodizer deduced from the measurements compared to the theoretical one. The measurement for the two first filters is biased by a too faint flux at the detector level. 
To obtain the theoretical curve, we made the calculation of the transmission coefficient of an Inconel coating on a BK7 glass substrate using thin film theory equations described by Born \& Wolf (1979~\cite{BornWolf79}) and the value of the Inconel refractive index determined by Goodell et al. (1973~\cite{Goodell73}).

Except for the four first filters, the measurement is totally in adequation with the numerical simulation.
This result validates the method used to calculate the theoretical curve.
We can conclude that on the spectral domain [450 nm - 700 nm], the transmission coefficient of the apodizer worths $39~\pm 3\%$.

\subsection{Apodizer global reflectivity in the visible}
\label{ssec:}
\begin{figure} [!h]
   \begin{center}   
 \includegraphics[height=5cm]{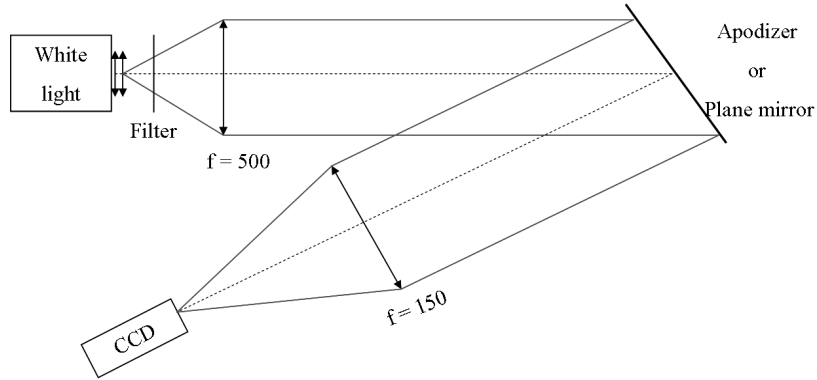}  
  \caption{Optical setup for the measurement of the apodizer reflectivity.}
  \label{fig:Schema_Principe_Mesure_Refl}
	\end{center}
   \end{figure}

 Figure~\ref{fig:Schema_Principe_Mesure_Refl} shows the optical setup to measure the apodizer reflectivity.

The apodizer is slightly inclined from the optical axis so as to  generate a reflected beam sufficiently disjoined from the incident one to be imaged on the CCD camera.
To obtain a reference measurement, the apodizer is replaced by a plane mirror in aluminium. The reflectivity coefficient of this mirror worths 90\%.
The measurements were carried out using the same experimental protocol as for the transmission coefficient measurement.
\begin{figure} [!h]
   \begin{center}   
 \includegraphics[height=5cm]{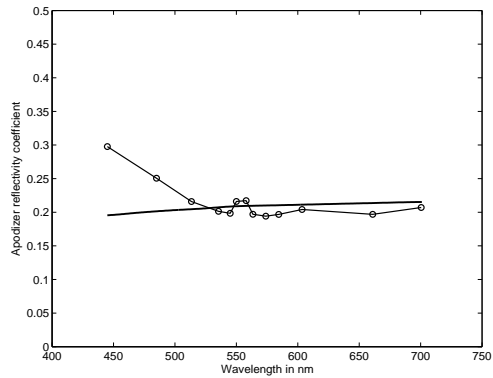}   
  \caption{Measurement of the apodizer reflectivity coefficient (dotted) compared with the numerical simulation (solid line).}
  \label{fig:Refl_Apo_3Lsd}
	\end{center}
   \end{figure}

Figure~\ref{fig:Refl_Apo_3Lsd} shows the evolution of the measured reflectivity coefficient compared, as previously, with the simulation obtained with the same computation code.
It wasn't possible to obtain exploitable data for filters number 1 and 2, the recorded flux was too faint.

\par\noindent Except for the 2 first wavelengths, the mesured curve complies with the numerical simulation.
The apodizer global reflectivity worths $21~\pm 1\%$ on the spectral domain [450 nm - 700 nm].

\section{ALC NULLING PERFORMANCES} 
\label{sec:Perf_nulling}

\subsection{Measurement principle}

Acquisitions are done in the final coronagraphic focal plane with the fibered laser diode ($\lambda = 633 nm$).
Three different Point Spread Functions (PSF) are successively recorded\,: the reference one, the reference apodized one and the coronagraphic one. It was necessary to place neutral densities behind the light source to avoid the saturation of the detector.
Thousand short exposures (with dark substraction) are co-added to obtain the long exposure.

\subsection{Results}

\begin{figure} [!h]
   \begin{center}  
(a) \includegraphics[width=0.26\linewidth]{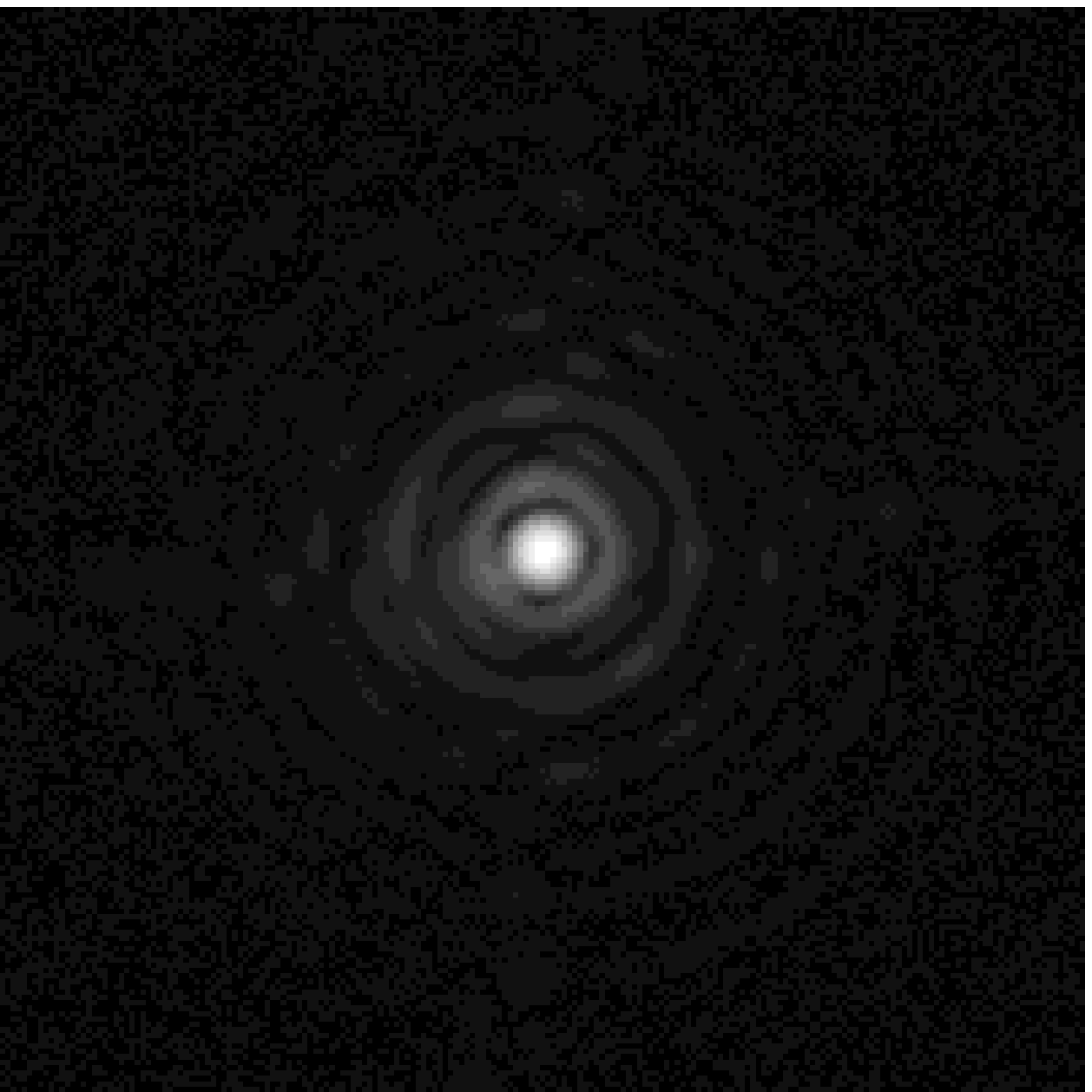}  \hfill
(b) \includegraphics[width=0.26\linewidth]{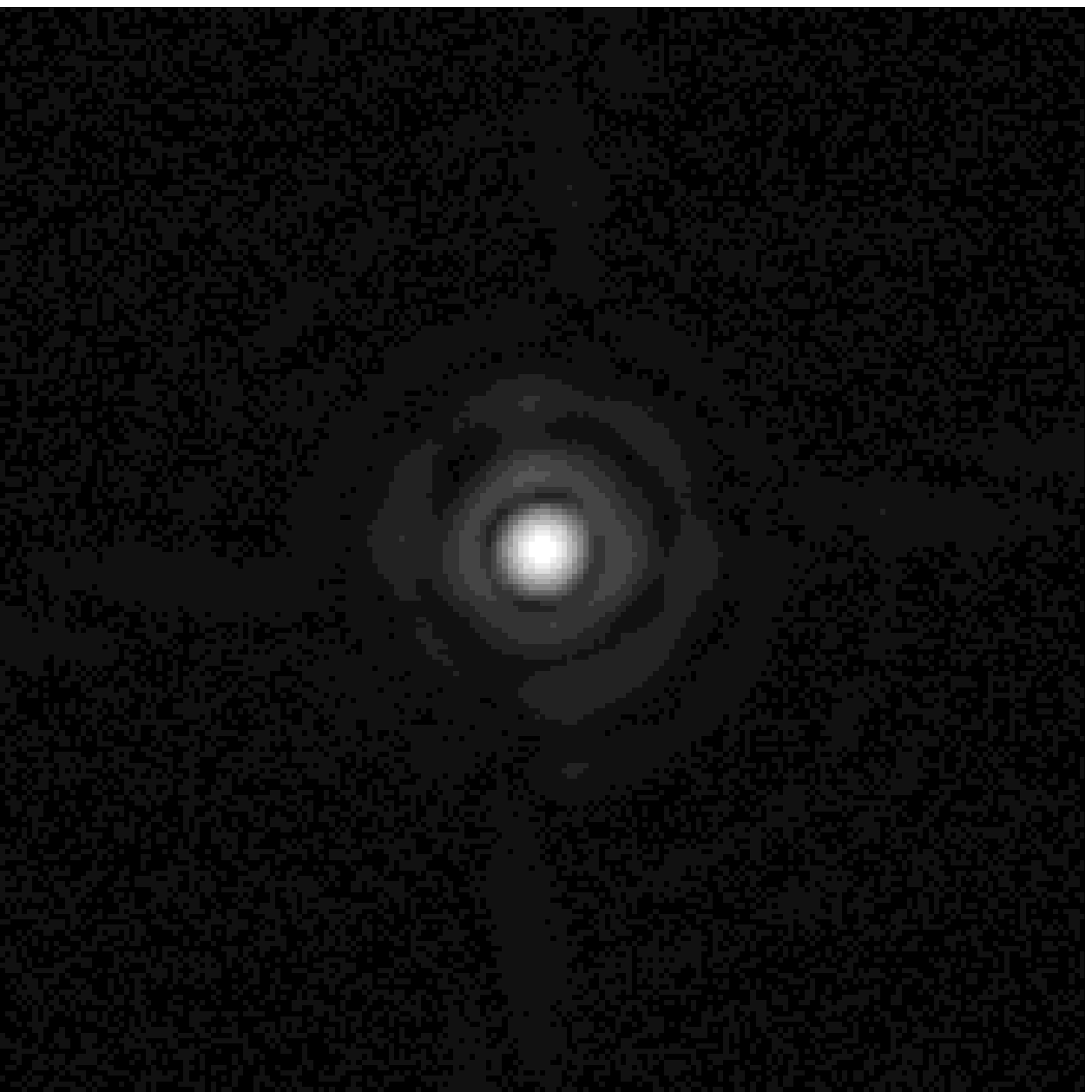}  \hfill
(c) \includegraphics[width=0.26\linewidth]{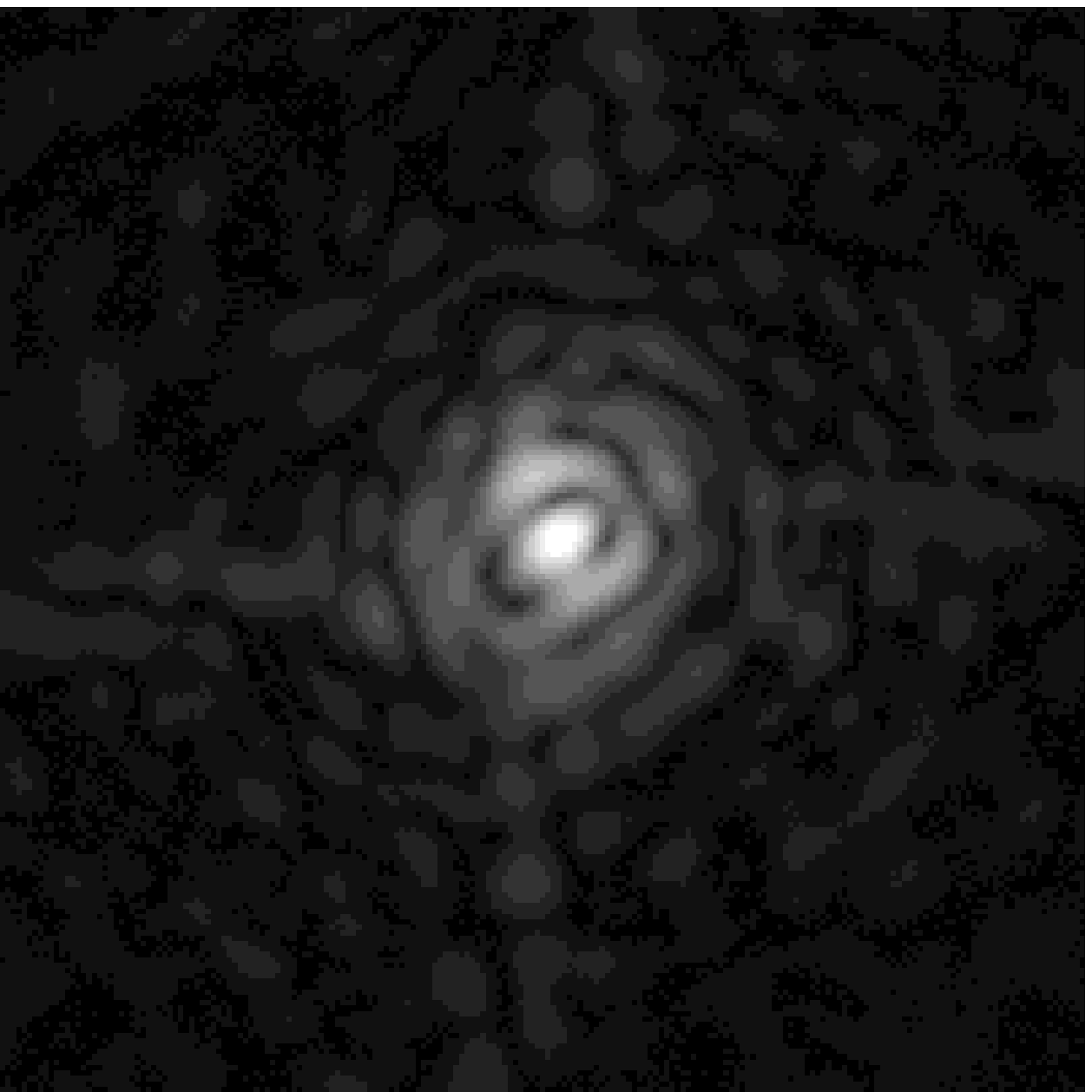}  \\
(d) \includegraphics[width=0.56\linewidth]{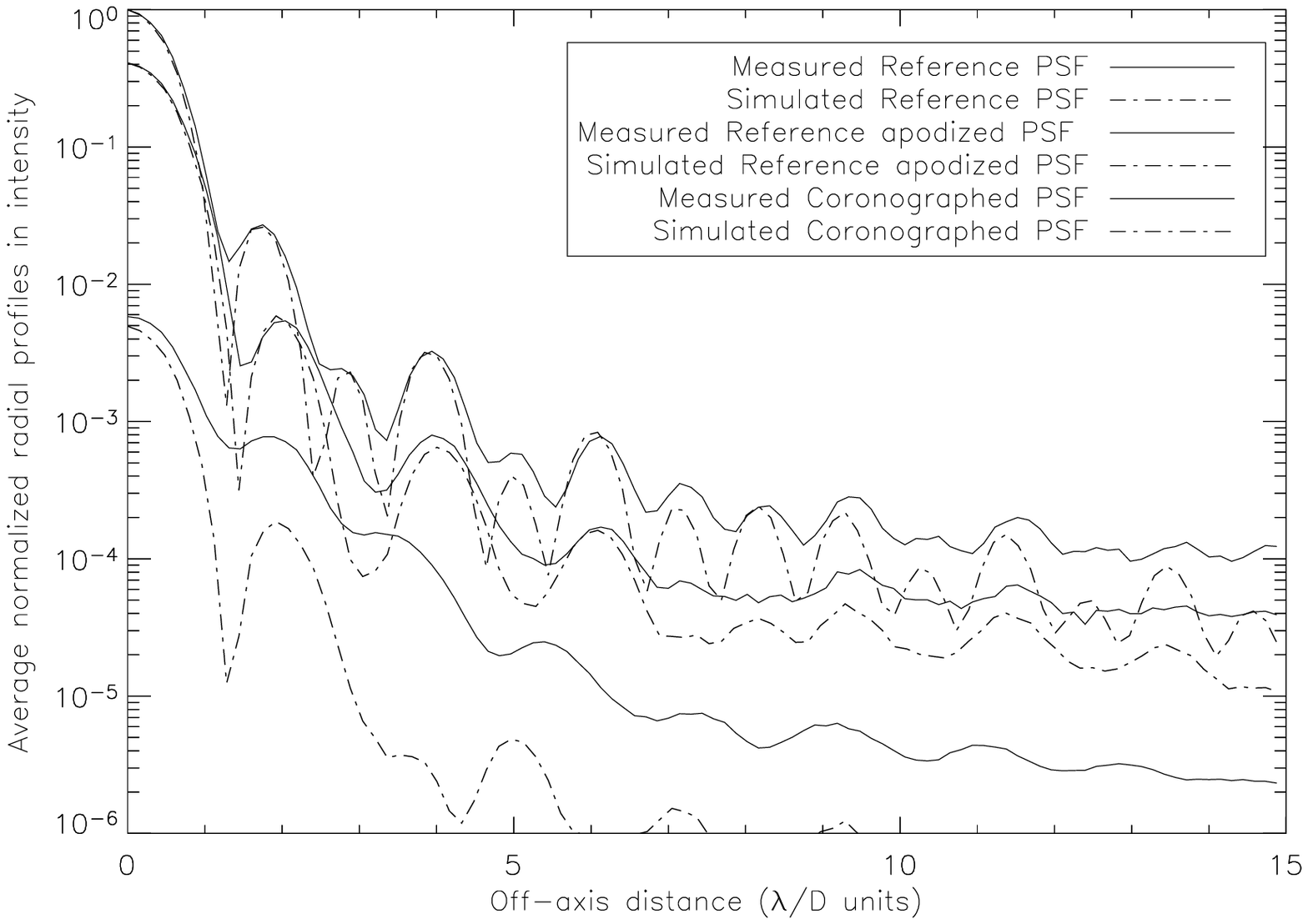}  
  \caption{Laboratory measurement at $\lambda = 633 nm$\,: (a) Reference non apodised PSF, (b) Reference apodised  PSF, (c) Coronagraphed apodized PSF, (d) average normalized radial profiles in intensity compared with simulations.}
  \label{fig:Mes_Compa_CLC_ALC}
	\end{center}
   \end{figure}

Fig.~\ref{fig:Mes_Compa_CLC_ALC} shows the measurements of the reference non apodized PSF (integration time $Ti=5~ms$, neutral density $ND=6.5$), the Reynard apodized PSF ($Ti=30~ms$, $ND=6.5$) and the coronagraphed PSF ($Ti=30~ms$, $ND=4.6$), compared with simulations. The latter takes into account the real transmission profile of the Reynard apodizer, as measured in Sect.~\ref{ssec:Trans_Apo_HeNe}.
For the 2D-images, the color scale was modified to enhance contrast and visibility.
The effect of the apodizer is clearly visible through the shift of the Airy rings betwen the reference and the apodized PSFs.
The coronagraphed apodized PSF has an elliptical shape. This effect, missing without the apodizer, is probably induced by the wavefront error or the coating defects introduced by the apodizer. 
The global shape of the measured PSFs is consistent with the simulations.
For off-axis distances greater than $\lambda/D$, there is an offset in intensity between experimental results and simulations in the PSF wings for both types of reference PSF.
In addition, the minimal intensity in the Airy rings of all the PSFs are much lower than that expected\,: it is essentially due to the fact that we are limited by a low dynamics of the 10-bits CCD camera.

Several sets of acquisitions were done\,: Table~\ref{tab:Perf_CLC_ALC_3Lsd} gives the ALC coronagraphic performances deduced from these measurements compared with simulations. For the peak attenuation, they agree with expectations. For the rejection ratio, the difference is due to the CCD low dynamics.

\begin{table} [!h]
\begin{center}
\begin{tabular}{|c|c|c|}
                                      								&   Experimental value          & Theoretical value  \\
\hline \hline
Peak attenuation for the ALC (from the apodized reference PSF)                 &       $71 \pm 11$              &  72                 \\
Peak attenuation for the ALC (from the reference PSF)           &     $170    \pm 25$                 &   202               \\
\hline\hline
Rejection ratio for the ALC (from the apodized reference PSF)                  &      $34  \pm 10$                 &    65               \\
Rejection ratio for the ALC (from the reference PSF)             &      $80  \pm 30$                &     157            \\
\hline
\end{tabular}
\end{center}
\caption{ALC coronagraphic performances at $\lambda = 633 nm$\,: comparison between experiment and simulations.}
\label{tab:Perf_CLC_ALC_3Lsd}
\end{table}
%

\newpage
\section{ALC SENSITIVITY TO MISALIGNMENTS OF ITS COMPONENTS} 
\label{sec:}
In this section, the 3 main components of the ALC, the apodizer, the coronagraphic mask, and the Lyot stop, are successively laterally then longitudinally misaligned from their ideal position. All measurements were done in the HeNe monochromatic light.

\begin{figure}[!h]
\begin{center}
Apodizer lateral and longitudinal misalignments\\  
\hspace{0.01cm}  \\
\includegraphics[width=0.46\linewidth]{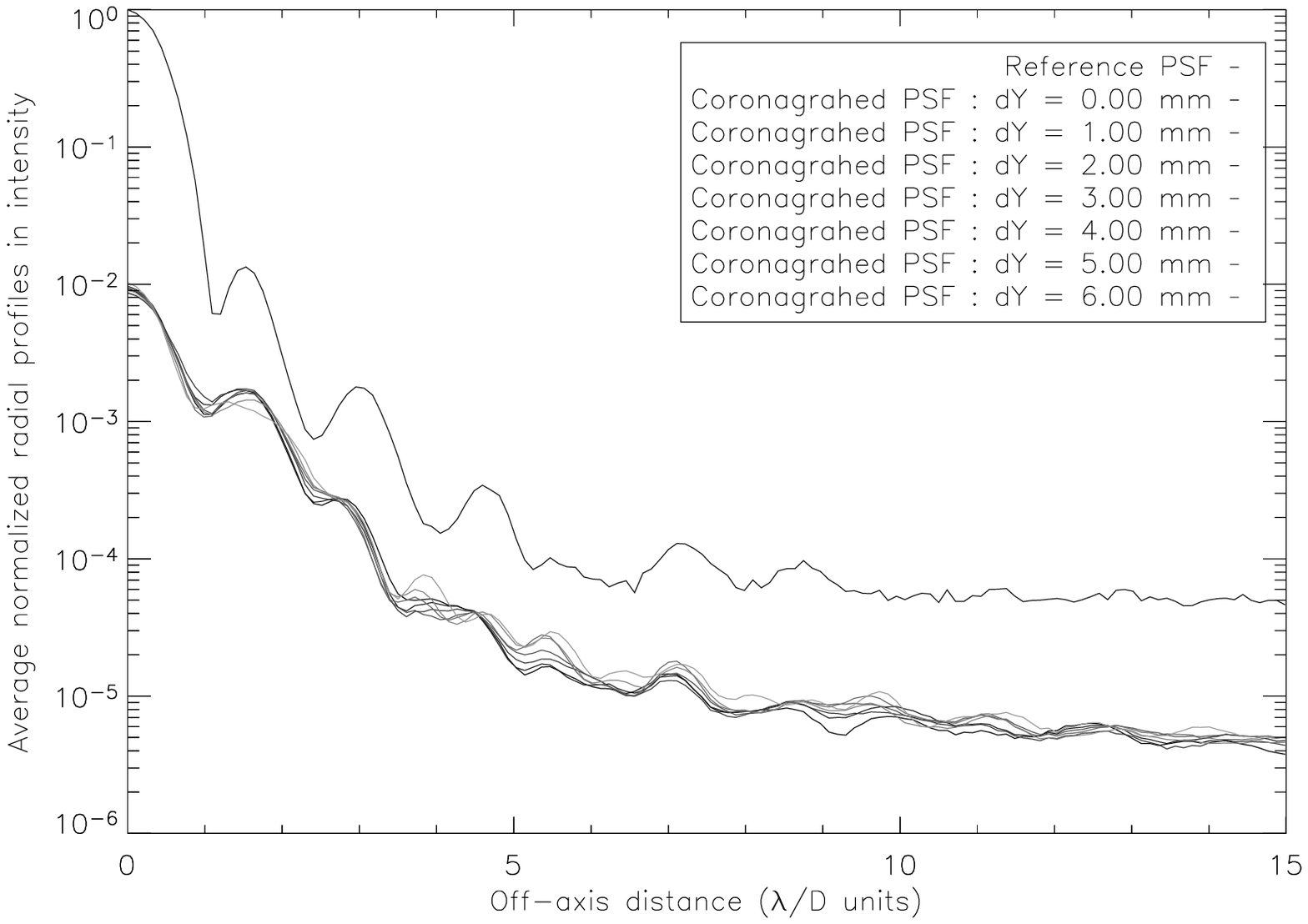}\hfill
\includegraphics[width=0.46\linewidth]{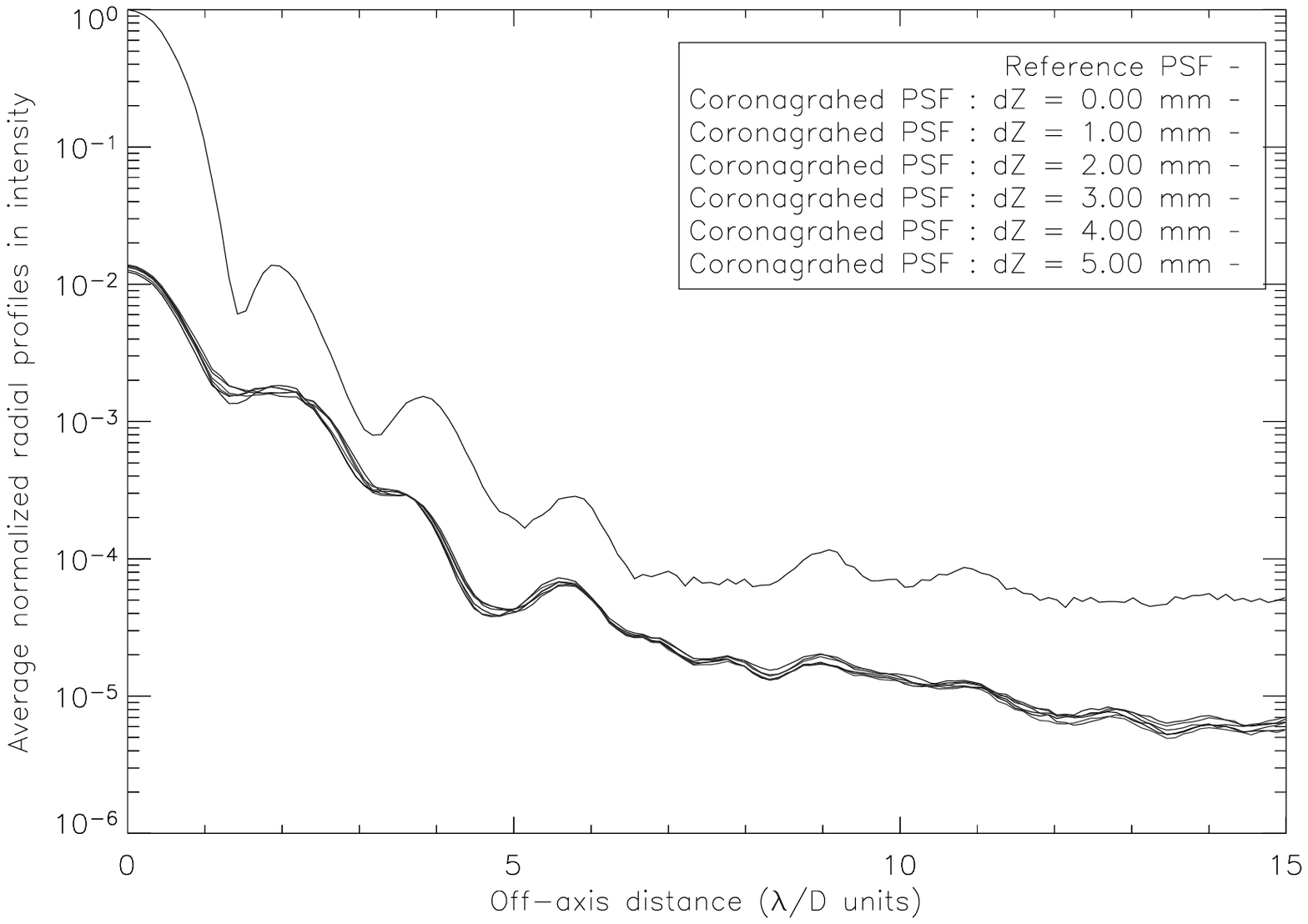}\\
\hspace{0.01cm}  \\
Coronagraphic mask lateral and longitudinal misalignments\\ 
\hspace{0.01cm}  \\
\includegraphics[width=0.46\linewidth]{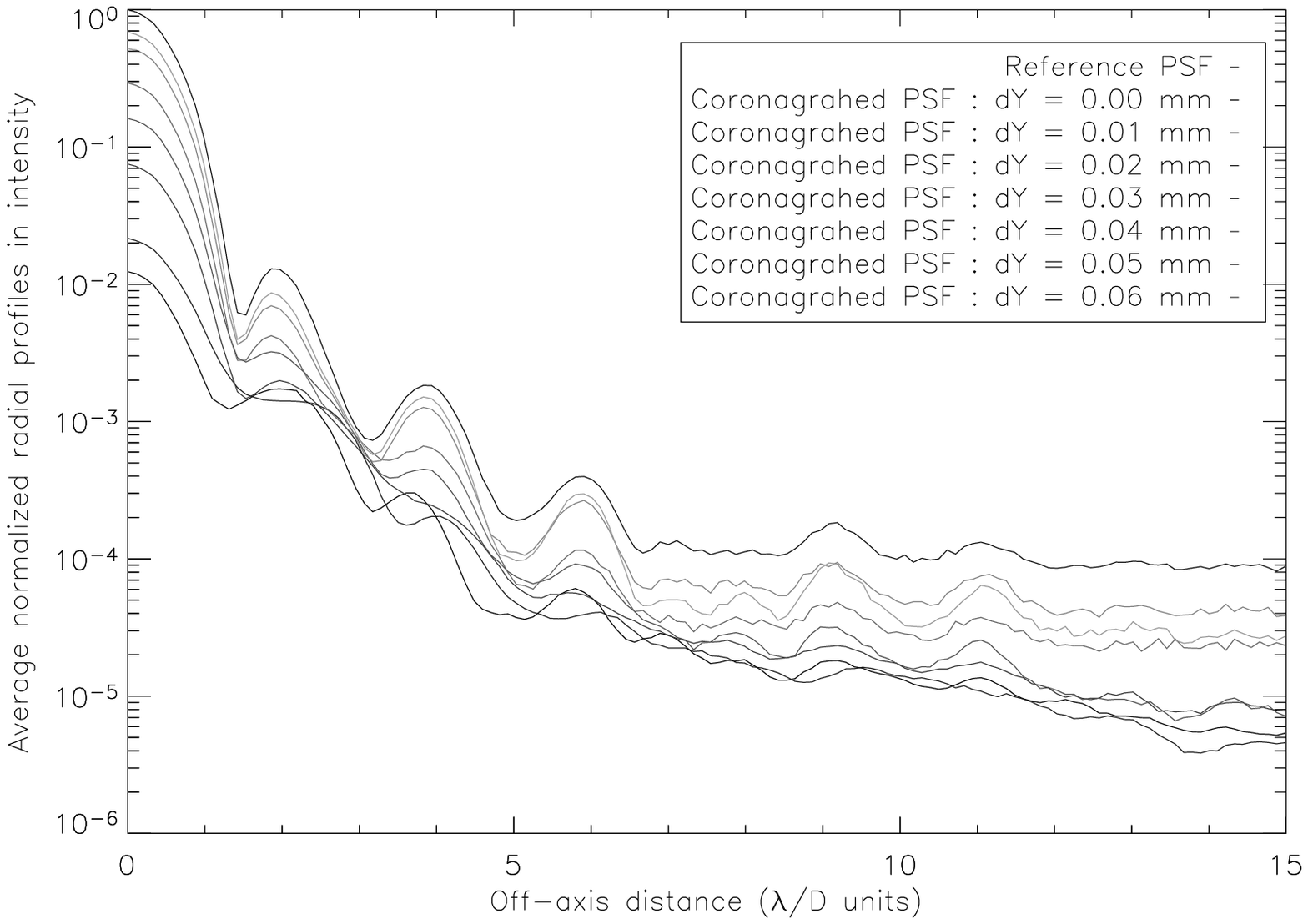}\hfill
\includegraphics[width=0.46\linewidth]{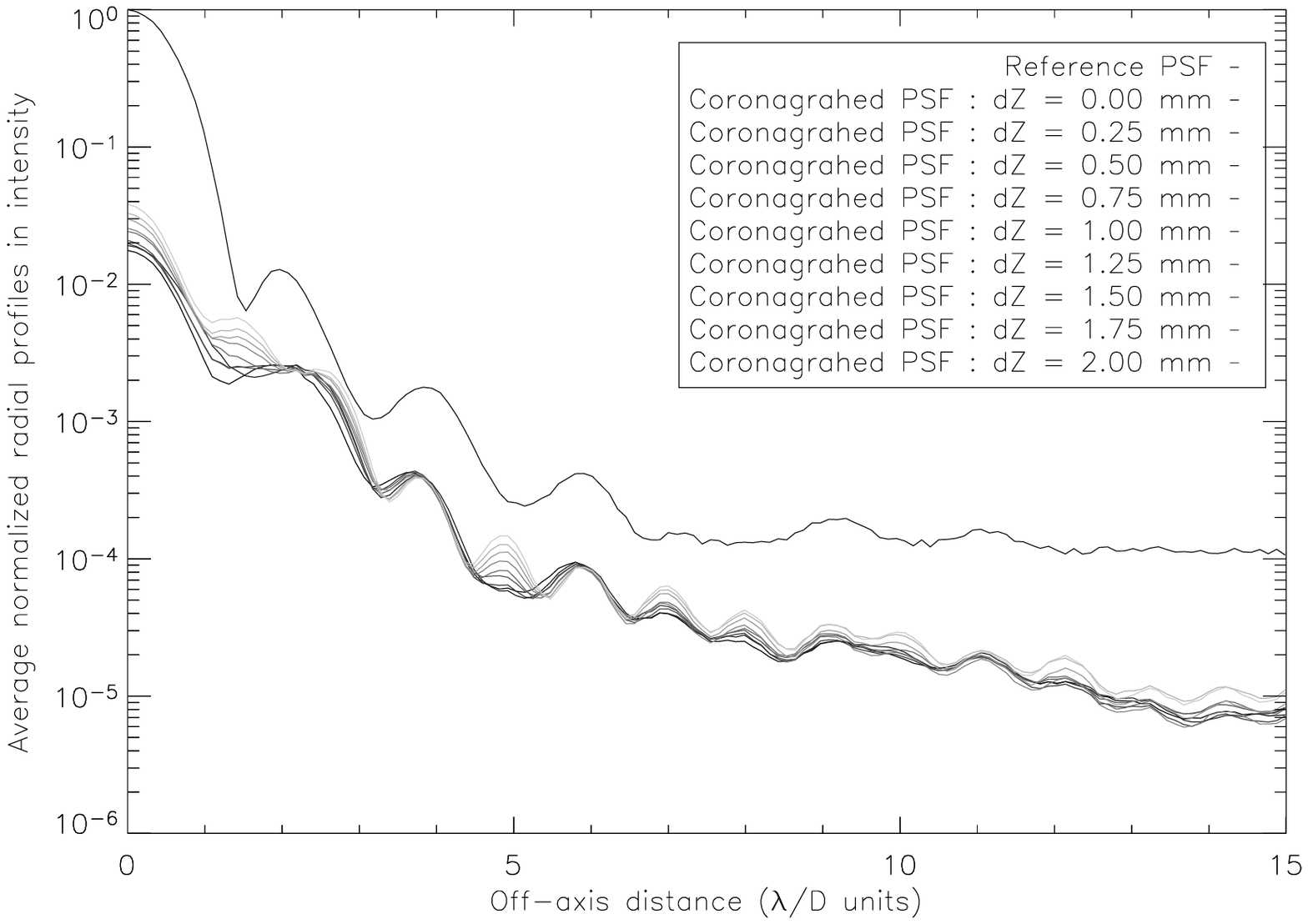}\\
\hspace{0.01cm}  \\
Lyot stop lateral and longitudinal misalignments\\ 
\hspace{0.01cm}  \\
\includegraphics[width=0.46\linewidth]{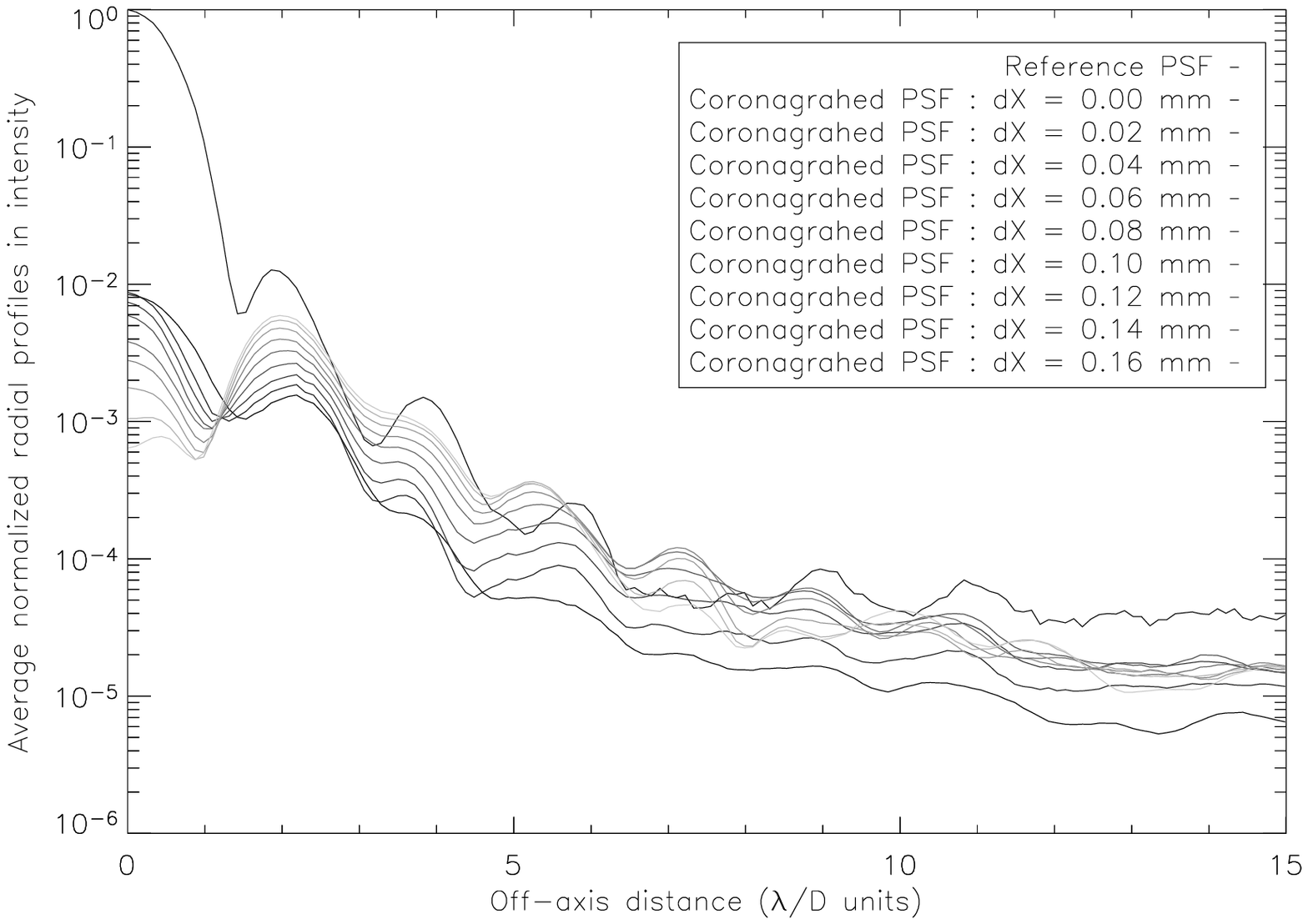}\hfill
\includegraphics[width=0.46\linewidth]{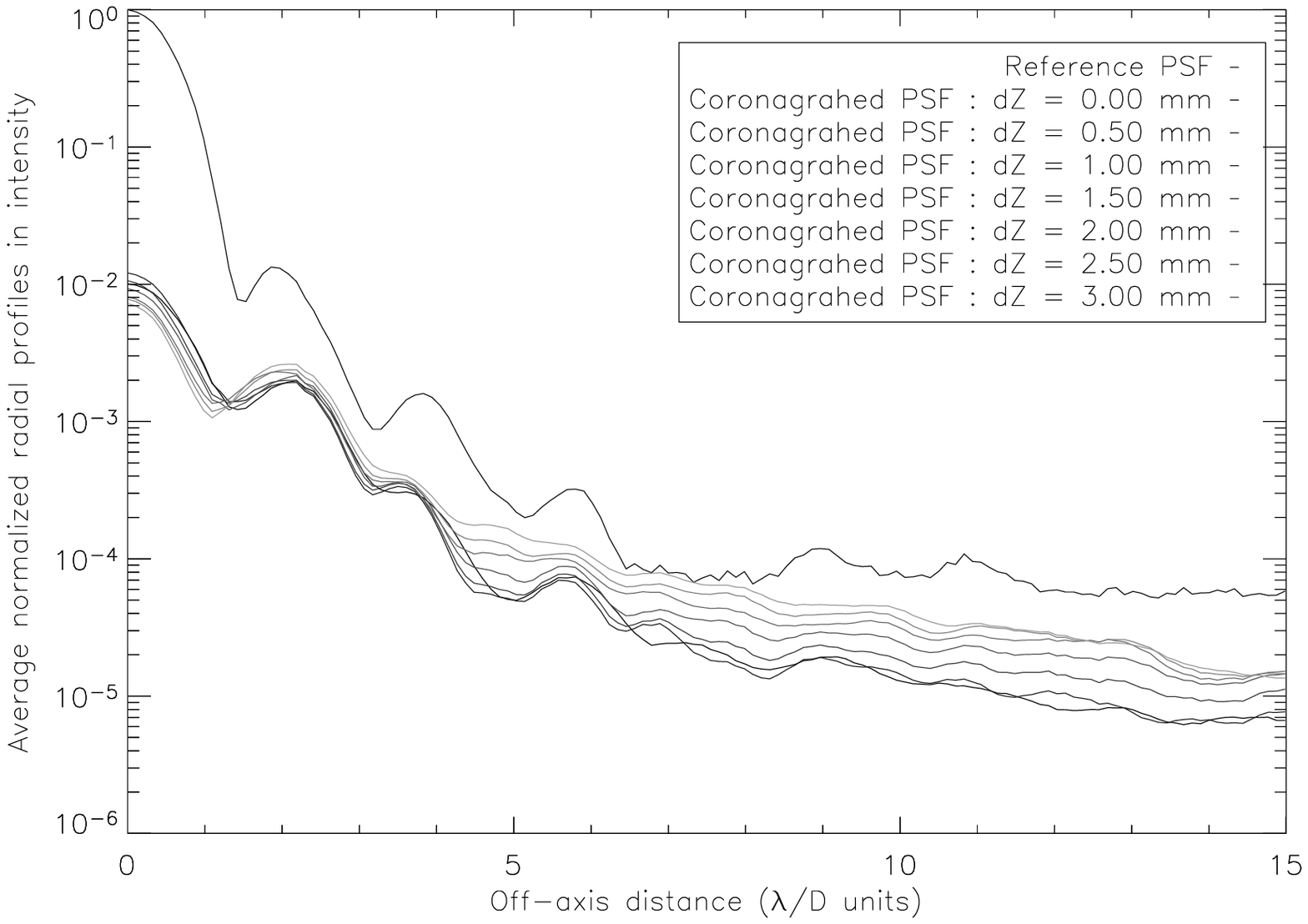}\\
\end{center}
\caption{Effect of the lateral and longitudinal displacements of the ALC main components\,: apodizer, coronagraphic mask and Lyot stop. For each component, the left curve corresponds to the lateral misalignment and the right curve to the defocus.}
\label{fig:All_Depl}
\end{figure}

Figure~\ref{fig:All_Depl} shows the resulting coronagraphed PSFs corresponding to both lateral and longitudinal misalignments for the apodizer, the coronagraphic mask and the Lyot stop.
We can identify with this figure what are the critical elements concerning the mechanical positionning on the bench\,: the coronagraphic mask in the lateral position and the Lyot stop in both positions.

\begin{figure}[!b]
\begin{center}
\includegraphics[height=6cm]{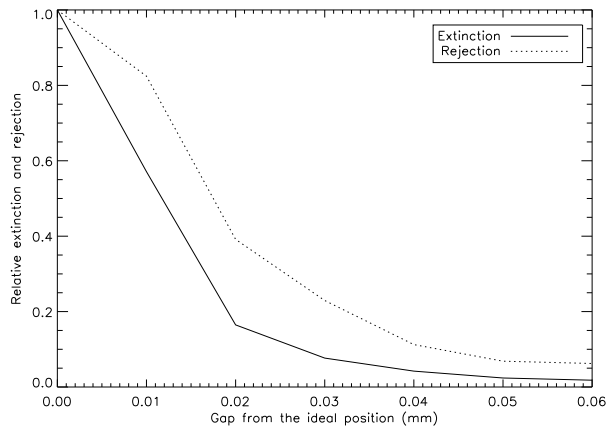}
\caption{Example of coronagraphic metrics used to determine the specifications on the positionning of the ALC components\,:
evolution of the extinction (solid line) and rejection (dash-dotted line) ratios as a function of the coronagraphic lateral displacement values $\Delta$Y, from 0 to 0.06~mm.}
\label{fig:expl_crit}
\end{center}
\end{figure}

In order to evaluate the specifications on the positionning of the ALC components, several coronagraphic metrics were used in addition to the
evolution of the PSFs shapes which is a subjective criterion.
We plotted the evolution of the extinction and rejection ratios as a function of the displacement value\,: an example concerning the coronagraphic
mask lateral misalignment is given on Figure~\ref{fig:expl_crit} and constitutes a good proof of the great dependency of the ALC performances to the lateral misalignment of the coronagraphic mask.
For each misalignment value, we also determined the contrast evolution, ie the ratio between the average radial profile of the coronagraphed PSF with the displacement, and the average radial profile of the coronagraphed PSF for the ideal position\,: we look for the displacement value inducing a loss of contrast lesser than 10\% that corresponds to the average level of noise. 

Table~\ref{tab:Tol_Pos_ALC} gives the specifications on absolute positioning for the three ALC components deduced from these criteria and  metrics.

\begin{table} [!h]
\begin{center}
\begin{tabular}{|c|c|c|}
Component                                     				&   Lateral tolerance              &  Longitudinal tolerance   \\
\hline \hline
Apodizer   							                        &    $ \pm~5~mm$                &      $\ge~5~mm$              \\
\hline
Lyot coronagraphic mask                                &    $\le~0.01~mm$               &     $ \pm~1~mm$               \\
\hline
Lyot stop                                                       &        $ \pm~0.04~mm$        &     $ \pm~0.5~mm$              \\
\hline
\end{tabular}
\end{center}
\caption{Synthesis of the tolerances values on the positioning of the ALC components obtained by measurements..}
\label{tab:Tol_Pos_ALC}
\end{table}

Furthermore, we can notice that the same tendencies were found with an other ALC prototype tested in the near Infrared (Guerri et al. 2008~\cite{Guerri08} ,
Boccaletti et al. 2008~\cite{Boccaletti08}).  

%

\clearpage
\section{CONCLUSION AND FUTURE PROSPECTS} 
\label{sec:concl}

We presented the results of the experimental characterization in the visible of an Apodized Lyot Coronagraph prototype that was designed for a preliminary concept feasibility study for the VLT-SPHERE instrument. Several type of measurements were carried out, the main conclusions that can be drawn are\,:
\begin{itemize}
\item apodizer characterization\,:
the transmission profile is out of the tolerance limits between radius $7~mm$ and radius $12.8~mm$. 
Nevertheless, the apodizer global transmission and reflectivity coefficients in the visible worth respectively $39\%$ and $21\%$ which is consistent with the simulations.

\item ALC coronagraphic performances in the visible\,: the PSFs global behavior and the extinction ratio are consistent with the simulations.
However, the lack in dynamics of the detector limits the rejection and the dynamics of the PSFs.
\item estimation of the ALC sensitivity to the lateral and longitudinal misalignment of its three main components\,: the coronagraph 
is very sensitive to lateral displacements of the coronagraphic mask and of the Lyot stop, and to the Lyot stop defocus.
\end{itemize}
We can finally conclude that these tests lead to consider the ALC as potential coronagraph for the VLT-SPHERE instrument and allowed the start of the study and the development of an ALC in the near-infrared suitably designed and dedicated to SPHERE (Carbillet et al. in prep.~\cite{Carbillet08}, Guerri et al. in prep~\cite{Guerri08}).

The future prospects concerning the use of the High Dynamics range imaging bench of the laboratory is the integration of a deformable mirror to generate wavefront residual errors (Guerri et al.~\cite{Guerri05}). The goal is to study the effects of these errors on the ALC coronagraphic performances.

\acknowledgments     
 The experiments reported in this article has been supported by the district of Provence Alpes C\^{o}te d'Azur (PACA), CNRS and ASHRA.
G.~Guerri is grateful to CNRS, the PACA district and Sud-Est Optique de Précision (France) for supporting her PhD thesis.


\bibliography{report}   
\bibliographystyle{spiebib}   

\end{document}